    Here the paper follows

\documentstyle[preprint,aps]{revtex}

\draft
\preprint{IP/BBSR/94-27}
\begin{document}
\begin{titlepage}
\title{Effect Of Single Particle Hopping And Out Of\\
Plane Magnetic Impurity On Coupled Planar Superconductors}
\author{Manas Sardar^{1} and Debanand Sa^{2}\\
Institute of physics, Bhubaneswar-751 005, India}
\footnotetext[1]{e-mail:manas@iopb.ernet.in}
\footnotetext[2]{e-mail:debanand@iopb.ernet.in}
\maketitle

\begin{abstract}
It is shown that the single particle band motion along the $c$
axis is harmful for superconductivity. Variation of $T_c$ with
$c$ axis hopping parameter is shown for both the conventional planar
superconductors and for interlayer pair tunneling mechanism of
Wheatley Hsu and Anderson(WHA).Effect of out of plane magnetic impurity
substitution is shown to suppres $T_c$ more for conventional
superconductors whereas there is very sharp decrease of $T_c$
in the WHA mechanism at larger concentrations.
\end{abstract}

\end{titlepage}

  All normal state properties of the high $T_c$ materials are
 highly anisotropic in nature. For example, resistivity anisotropy
$\rho_{c}/\rho_{ab}$ is about $10^3$ to $10^5$ in Bi compounds\cite{Iye}.
On the other hand the typical anisotropies in the superconducting
phase like $\lambda_{ab}/\lambda_{c}$ and $\xi_{ab}/\xi_{c}$
are much smaller (of the order of 5-10). This shows that the
superconductivity
is a real 3 dimensional phenomena, with the coupling between the CuO
planes being a very relevent parameter. The normal to superconductor
transition is at the same time a  two to three dimensional transition.

          Most theories of high $T_c$ materials are purely
two dimensional in
 nature, where the coupling between the planes is ignored to begin with.
The large semiconducting type $c$ axis resistivity ( greater than
Mott limit al low temperatures ) is shown as a proof, that the
elctrons have no band motion along the $c$ axis\cite{ander}. In other words
$c$ axis motion is fully incoherent. Invoking localisation along
the $c$ axis is meaningless, because electrons cannot localise
along one direction only\cite{ander}. It has been emphasized by
Anderson\cite{ander}, that the single particle band term along
the $c$ axis is inoperative in the normal state, and in the
superconducting state as well.
On the other hand recently it has been argued by
Rojo et. al, \cite{lev} that the large
c-axis resistivity is not inconsistent with a finite hopping
amplitude between the planes, because the off-diagonal disorder
has a delocalization effect.
For the superconducting state, at a phenomenological level they are described
by a Lorence-Doniach kind of
model\cite{Bul}. Here the two adjacent CuO layers ( who are
individually superconducting
, coming from any of the existing purely 2-d mechanisms) have a Josephson
coupling between them. This coupling further enhances the
transition temperature
of the individual layers. A Josephson coupling between the planes tunnels
pairs of electrons between the planes. One starts from an effective
BCS hamiltonian for the two planes, and switches on a single particle hopping
term in the c-direction. Josephson coupling occurs between the
planes in second
order of this single particle hopping amplitude.

   We explicitly show here that, having a single particle tunneling
term in the c-direction is harmful for superconductivity. This is because,
as far as one plane is concerned, this acts as a pair breaking
perturbation. So even though, the Josephson coupling  leads to real
3-d coherence and an apparent increase in transition temperature, the
single particle hopping between the planes ($t_{\perp}$) tries to destory
superconductivity, and the $T_c$ is very much suppresed compared to the
purely decoupled 2-d superconductors, because the single particle and
pair tunneling have opposite effects on $T_c$. We find that, for the model
where, two planar BCS superconductors are coupled by both single particle
and josephson tunneling, the $T_c$ decreases with increase of $t_{\perp}$
slowly at first and very steeply at larger values. It is a monotonous
decrease of $T_c$ , in other words,single particle tunneling and consequent
reduction of $T_c$ due to pair breaking always plays a dominant role.

    On the other hand , in the interlayer pair tunneling mechanism
of Wheatley , Hsu and Anderson \cite{whe}(WHA), it is argued that in
the normal state
there is no band motion of electrons in the c-direction, even though the
hopping amplitude $t_{\perp}$ is quite substantial as many band theory
calculations shows. This is so, because of the underlying assumpsion
of spin-charge decoupling of the electronic system in the
2-d plane due to strong correlation. Therefore, even though $t_{\perp}$ is
quite large , it is not effective in tunneling electrons in the c-direction
simply because there are no low energy electron like quasiparticle
near the Fermi surface in the 2-d plane ( c-direction conduction is
supposed to be purely incoherent in nature). In this mechanism, it is
proposed that even in the superconducting phase single particle band motion
is absent. The first channel of c- axis conductivity occurs in the second
order in $t_{\perp}$, that is through Josephson pair tunneling.
Incoherent motion of single electrons , but coherent tunneling of
pairs of electrons is shown to be possible in model hamiltonian
by Muthukumar et al\cite{muthu}
Here $T_c$ increases with increase in $t_{\perp}^2$ unlike in the
earlier case where $T_c$ decreases with increase in $t_{\perp}$.

  Next we consider the effect of magnetic impurity substitution out
of the plane . There is a dramatic suppression of $T_c$ upon
substituting Y by Pr in YBCO compound , where Pr ions show a net
magnetic moment $(\approx 2.7\mu_B)$ as has been observed in the
high temperature susceptibility data \cite{sod}. We consider
the case , where out of plane magnetic impurity
have no direct exchange coupling ( of the local Kondo kind) with the
conduction electrons in the plane. Also there  is no hybridization of the
impurity levels with the O or Cu orbitals. In other words , the presence
of the moment does not change the in plane electronic
parameters. In contrast, Fehrenbacher et. al, \cite{ric} has proposed
that Pr electronic levels hybridize with the planar Oxygen
leading to a decrease in the inplane hopping amplitude.

In the present situation, we show that for more conventional theories , where
single particle motion in the c-direction is operative, there will be,
(1) strong suppresion of $T_c$ due to
spin flip scattering by the impurity moment
with the electrons moving along c-axis.
(2) The second channel of conduction along the c-axis, that is the
pair tunneling process, will also be affected by the magnetic impurity.
The effect can be modelled , as if the Cooper
pairs get a phase slip of $\pi$ while travelling through the
impurity center \cite{shi}.
This will reduce effective pairing potential and hence reduce $T_c$.
We will show that the first process of reduction of $T_c$ is
more dominant than the second one, because for moderate
values of $t_\perp$, the Josephson
tunneling between the planes, even in the absence of impurity
increases $T_c$ very slowly with increase of $t_\perp$.
However, in the Wheatley Hsu Anderson mechanism(WHA) , the
single particle tunneling
is absent. Only the interlayer pair tunneling , will be affected by
the presence of the moments.
In the WHA mechanism, the pair tunneling term is peculiar, in the sense
that, in the process of pair tunneling, the individual momenta of the
partners of the cooper pairs are conserved. So the pairing term in the
hamiltonian there is only one momentum sum rather that the conventional
two momenta sum.
The pairing potential is extremely local in momentum space.
This has a remarkable effect on $T_c$. The $T_c$
increases with increase of pair tunneling amplitude( which is
quadratic in $t_\perp$) much more steeply compared to the
usual Josephson coupling case. Theoretically it is argued that, the
pequliar momenta conserving pair tunneling is a consequence
of the normal state being a Luttinger liquid\cite{ander}.
This Josephson coupling will decrease with increase of magnetic
impurity concentration due to phase slippage leading to decrease of $T_c$.
We find that for low impurity concentration the $T_c$ falls faster
with impurity concentration in the conventional planar models, but
at larger concentrations ,$T_c$
falls faster in the WHA mechanism.

To begin we consider the hamiltonian,

\begin{eqnarray}
H~~=~~\sum_{k}((\epsilon_{k}-\mu)c_{k\sigma}^{1\dagger}c_{k\sigma}^{1}
+ 1\rightarrow 2) +t_{\perp}\sum_{k}(c_{k\sigma}^{1\dagger}c_{k\sigma}^{2}
+  h.c.)
\nonumber \\
{}~+~\sum_{kk^{\prime}}(V_{kk^{\prime}}
c_{k\alpha}^{1\dagger}c_{-k\beta}^{1\dagger}
c_{-k^{\prime}\beta}^{1}c_{k^{\prime}\alpha}^{1}+1\rightarrow 2)
+{t_{\perp}^{2}\over t}\sum_{kk^{\prime}}(c_{k\alpha}^{2\dagger}c_{-k\beta}
^{2\dagger}c_{-k^{\prime}\beta}^{1}c_{k^{\prime}\alpha}^{1} +  1\rightarrow 2)
\end{eqnarray}

  Here all momenta are 2-d momenta. We consider a 2 layer per unit cell
material. $c_{k}^1$ and $c_{k}^2$ are electron annihilation operators
in layer 1 and 2. $\epsilon_{k}$ is the free dispersion in the plane and
$t_{\perp}$ is the c-axis hopping amplitude. $V_{kk^{\prime}}$
is a BCS type pairing potential in the plane, coming from
any conventional mechanism,
details of which are of no consequence for our purpose. ${t_{\perp}^2\over
t}$ is the Josephson coupling term. We have not taken any momentum dependence
of the hopping amplitude $t_{\perp}$ along the $c$ axis.
$V_{kk^{\prime}}$ is assumed to have the form,

\begin{flushleft}
\begin{eqnarray}
&&V_{kk^{\prime}}~~=
\left \{
\begin{array}{ll}
{}~~-V~~~,~~~ {\rm for}~~~ \epsilon_{F}-\hbar \omega_{c}\langle
\vert \epsilon_{k}\vert,\vert \epsilon_{k^{\prime}}\vert \langle \epsilon_{
F} +\hbar \omega_{c} \\
{}~~~~~ 0~~~,~~~ {\rm otherwise}
\end{array}
\right.
\end{eqnarray}
\end{flushleft}

\noindent Where $\hbar \omega_{c}$ is the cutoff energy coming
from a more microscopic origin.
We assume that the in-plane pairing interaction comes from electron phonon
interaction. So $\omega_c$ will be the Debye frequency.
For simplicity we assume that there is only one cutoff in the theory
set by the in plane BCS coupling, and Josephson coupling also operates within
the same cutoff.

Now we do the mean field, by putting in the pairing ansatz
$$
\langle c_{k\uparrow}^{1\dagger}c_{-k\downarrow}^{1\dagger}\rangle~~=
\langle c_{k\uparrow}^{2\dagger}c_{-k\downarrow}^{2\dagger}\rangle~~=
\bigtriangleup^{\star}
$$
Then the third and fourth term can be combined into
$$
(V+{t_{\perp}^{2}\over t})\sum_{k}^{\prime} (\bigtriangleup^{\star}
c_{k\downarrow}^1c_{k\uparrow}^1  +\bigtriangleup c_{k\uparrow}^{1\dagger}
c_{-k\downarrow}^{1\dagger} +1\rightarrow 2)
$$
Where the prime over the summation represents a restricted sum
within the Debye Cut-off. To take into account the single
particle hopping between the planes, we define two kinds of fermions
$$
c_{k\sigma}^1=~~{1\over 2}(\phi_{k\sigma} +\psi_{k\sigma})~~ {\rm and}~~
c_{k\sigma}^2=~~{1\over 2}(\phi_{k\sigma} -\psi_{k\sigma})
$$
In terms of them the mean field hamiltonian will be

\begin{equation}
\sum_{k}(\epsilon_{k}-\mu +t_{\perp})\phi_{k\sigma}^{\dagger}\phi_{k\sigma} +
\sum_{k}(\epsilon_{k}-\mu -t_{\perp})\psi_{k\sigma}^{\dagger}\psi_{k\sigma} \\
+ (V+{t_{\perp}^2\over t})\sum_{k} [(\bigtriangleup^{\star} \phi_{-k\downarrow}
\phi_{k\uparrow} +
\bigtriangleup \phi_{k\uparrow}^{\dagger}
\phi_{-k\downarrow}^{\dagger} ) + \phi \rightarrow \psi]
\end{equation}

 $\phi$ and $\psi$ fermions describes the electrons in the bonding and
antibonding bands. The hamiltonian looks like a sum of two BCS reduced
hamiltonins for the bonding and antibonding electron systems. The generalised
gap equation will be

\begin{equation}
{1\over (V+{t_{\perp}^2\over t})}~~=~~ {1\over 2}\sum_{k} {tanh( \beta E_{k}
^{\phi}/2) \over 2 E_{k}^{\phi}} ~~+ ~~
{1\over 2}\sum_{k} {tanh( \beta E_{k}
^{\psi}/2) \over 2 E_{k}^{\psi}}
\end{equation}

where,
$$
E_{k}^{\phi ,\psi} ~~=~~\sqrt {(\epsilon _{k}\pm t_{\perp})^2 +
\bigtriangleup ^2 }
$$
Note that the summations over momenta in the first and second term
are over two different energy shells centered around $\mu \pm t_{\perp}$.
Going from summation to integral and converting to energy variables
it is not very difficult to see that the $T_c$ is given by

\begin{equation}
k_BT_c~~=~~\sqrt { \omega_{c}^2-t_{\perp}^2} {2e^{\gamma}\over \pi}
e^{- {1\over N(0) (V+t_{\perp}^2/t)}}
\end{equation}

for small values of $t_{\perp}$,
where $e^{\gamma} =1.781$. It is clear that the $T_c$ decreases with
increase in $t_{\perp}$ or more or less insensitive to it depending
on the magnitude of $\omega_c$ and the in plane BCS coupling. Major
effect of the out of the plane single particle hopping is to shrink
the cutoff of the effective BCS interaction potential.
Physically one should think of the single particle hopping in the
c-direction acting as a pair breaking mechanism, and thereby
destroying superconductivity. Energetically the condensation
energy lost by losing superconductivity can be compensated by the gain in
the single particle kinetic energy in the C direction. For
larger values of $t_{\perp}$ , band splitting will be larger and the chemical
potential will be very near the band edge of one of the subbands, for low
doping, while the other band will be submerged much below the Fermi surface.
This kind of scenario has been proposed by Levin and Quader\cite{levin} to
explain the transport properties in the normal state of the two layer
materials. We do not consider this limit.

      Next we consider the case where, there is some magnetic impurity in
      between the planes. Within the 1-band $t-J$ model scenario\cite{rice},
the tunneling
      process in the c-direction is a two
step process, where the inplane hole( Zhang
      -Rice singlet) moves over to the $Y$ 6s orbital( for 123 compound)
      and from there to the ZR
      singlet in the upper plane. If one substitutes the $Y$ atom by some other
      atom having a net magnetic moment, then this
will scatter the electrons moving
      along the c direction. One can model the effect of the impurity by the
      interaction hamiltonian,
     \begin{equation}
      H^{\prime}~~=~~U_{2}\vec S\cdot\vec \sigma
     \end{equation}
      Where $\vec S$ denotes the impurity spin and $\vec \sigma$ the electron
      spin density. This will flip the spin of the electrons travelling
      along the c direction. We can rewrite the above Hamiltonian in the form
      \begin{equation}
      H^{\prime }~~=~~\sum_{kk^{\prime}}U(k-k^{\prime})[(c_{k\uparrow}
      ^{2\dagger}c_{k^{\prime}\downarrow }^1 +
      c_{k\downarrow}
      ^{2\dagger}c_{k^{\prime}\uparrow }^1) + ~1\rightarrow 2]
      \end{equation}
      It is reasonable to assume that the scattering will be predominantly
      in the forward or backward direction only. Also since the translational
      symmetry is broken only in the c direction , the inplane momenta
      should be conserved. So the interaction hamiltonian will be,
      $$
     U(0) \sum_{k}(c_{k\uparrow}^{2\dagger} c_{k\downarrow}^1 +c_{k\downarrow}
      ^{2\dagger}c_{k\downarrow}^1 ~~+~~1\rightarrow 2)
      $$
      for the forward scattering and similarly, there will be backward
      scattering terms like,
      $$
      U(2k)\sum_{k}(c_{k\uparrow}^{1\dagger}
c_{k\downarrow}^1 +c_{k\downarrow}
      ^{1\dagger}c_{k\downarrow}^1 ~~+~~1\rightarrow 2)
      $$
       Going to the $\phi $ and $\psi$ fermion representation, we get
       $$

H^{\prime}~~=~~U_{eff}\sum_{k}(\phi_{k\uparrow}^{\dagger}\phi_{k\downarrow}
	+\phi_{k\downarrow}^{\dagger}\phi_{k\uparrow} + \phi \rightarrow \psi
	$$
Where $U_{eff}=U(0)+U(2k)$.
	In terms of these fermions, the interaction hamiltonian looks
	like {\it a direct exchange coupling of the impurity moment with
	the bonding and antibonding band electrons}.
	This will lead to a further reduction in the transition temperature
	as discussed by Maki\cite{maki}. The modified $T_c$ will be,
	$T_c =T_{c0}-{\pi\over4}{1\over \tau_{2}}$, where
	$$
	{1\over \tau_{2}}~~=~~2\pi U_{eff}^2 n N(0)S(S+1)
	$$
	where $n$, $N(0)$ and $S$ are the impurity density, conduction
	electron density of states near the Fermi surface and the impurity
	magnetic momont respectively. As discussed by Maki, not only the
	$T_c$ will be suppressed and superconductivity destroyed beyond a
	certain concentration of impurity, but also , for moderate density of
	moments one will see a finite density of states within the gap.
	This could be observed in the tunneling and photoemission
	experiments.

  Moving over to the pair tunneling from layer to layer, the tunneling hamil
tonian can be written in the form,
$$
\sum_{kk^{\prime}}t_{\perp kk^{\prime}}\delta_{ss^{\prime}} ~~+~~
U_{kk^{\prime}}\sigma_{ss^{\prime}}S_i )c_{k\sigma}^{1\dagger}
c_{k^{\prime}\sigma^{\prime}}^{2}~~+~~h.c
$$
$S_i$ is the operator for the local moment at site $i$ and $\sigma$ s are the
Pauli matrices. It is not difficult to see that , whenever
the cooper pair encounters a magnetic moment while
travelling along the $C$ axis, the corresponding
pair tunneling amplitude gets reduced from ${t_{\perp}\over t}$ to
${(t_{\perp}^2-U_{eff} s(s+1))\over t}$.
The mean field pair tunneling hamiltonian will be after impurity average,
$$
{(t_{\perp}^2-nU^2_{eff}s(s+1))\over t}\sum_{k} (\bigtriangleup^*
\phi_{-k\downarrow}
\phi_{k\uparrow} ~~+~~\bigtriangleup \phi_{k\uparrow}^{\dagger}
\phi_{-k\downarrow}^{\dagger} ~~+~~\phi \rightarrow \psi
$$
where $n$ is the concentration of magnetic impurity substituted in
between the planes.
Corresponding gap equation will be,
$$
{1\over V_{eff}}~~=~~{1\over 2}\sum_k {tanh(\beta E_k^{\phi /2})\over
2 E_k^{\phi}} ~~+~~
{1\over 2}\sum_k {tanh(\beta E_k^{\psi /2})\over
2 E_k^{\psi}} ~~+~~
$$
where $V_{eff}=V+t_{\perp}^2/t -n U^2_{eff}s(s+1)/t$
, remembering of course that with the introduction of single
particle hopping the $T_c$ will be further reduced the way we
indicated before.

In the case of WHA mechanism, modified recently by
Chakraborty et al\cite{cha},
The full hamiltonian in absence of impurity is,
\begin{eqnarray}
\sum_k (\epsilon_k-\mu)c_{k\sigma}^{1\dagger}c_{k\sigma}^1 +1\rightarrow 2
{}~~+~~V\sum_{kk^{\prime}} c_{k\uparrow}^{1\dagger}c_{-k\downarrow}^{1\dagger}
c_{-k^{\prime}\downarrow}^2c_{k^{\prime}\uparrow}^2 ~~+~~h.c
{}~~+1\rightarrow 2 \\
{}~~+~~
t_{\perp}^2/t\sum_k c_{k\uparrow}^{1\dagger}c_{-k\downarrow}^{1\dagger}
c_{-k\downarrow}^2c_{k\uparrow}^2 ~~+~~h.c ~~~+1\rightarrow 2
\end{eqnarray}
Notice the difference in the Josephson tunneling term in Chakraborty
et al's hamiltonian from the conventional Josephson terms.
The gap equation will be,
$$
\bigtriangleup _k={t_{\perp}^2\over t} {\bigtriangleup_k\over 2E_k}tanh(\beta
E_k/2) ~~+~~V\sum_q {\bigtriangleup_q\over 2E_q} tanh(\beta E_q/2)
$$
where $E_q=\sqrt {E_q^2+\bigtriangleup_q^2}$
In the presence of magnetic impurity the gap equation will be modified to,
$$
\bigtriangleup _k={(t_{\perp}^2-nU^{2}_{eff}s(s+1))\over t}
{\bigtriangleup_k\over 2E_k}tanh(\beta
E_k/2) ~~+~~V\sum_q {\bigtriangleup_q\over 2E_q} tanh(\beta E_q/2)
$$
We have solved all the gap equations numerically to locate the $T_c$.

We take the
inplane dispersion to be $\epsilon_k=-2t({\rm cos}~k_x~+~{\rm cos}~k_y)
{}~+~4t^{\prime}{\rm cos}~k_x{\rm cos}~k_y~-\mu~$, with $t=0.3~{\rm eV}$,
 $t^{\prime}=0.1125~{\rm eV}$, $\mu=-0.45~{\rm eV}$, $v=0.27~{\rm eV}$
and $t_{\perp} =0.1~{\rm eV}$.
In Fig.1 we show the $T_c$ variation with $t_{\perp}$ for interlayer
tunneling mechanism and the usual Josephson coupled superconductors , with and
without the band term along $c$ axis. We find that,
(1) For the interlayer tunneling mechanism the $T_c$ rises with increase
in $t_{\perp}$ very steeply. For example, with $t_{\perp}=0.0$ we fix
$V=0.22$ to get a $T_c$ of 5 degrees. But for $t_{\perp}=0.1$ the
$T_c$ increase to 85 degrees.
(2) For the usual Josephson coupled superconductor without single particle
hopping, $T_c$ rises very slowly with $t_{\perp}$. $T_c$ is only
35 degrees for $t_{\perp}=0.1$.
(3) With single particle hopping term included, remarkably the $T_c$
decreases with increase in $t_{\perp}$. We emphasize that, there is no obvious
reason why the single particle hopping  along the $c$ axis
should be absent in conventional fermi liquid theories..
This is one of the important differences between the conventional
Josephson coupling and Anderson's Josephson term.

 In Fig.2 , we show the $T_c$ variation with the
pair breaking parameter $\alpha =
 U^2_{eff} ns(s+1)$, which is directly proportional to the
 magnetic impurity concentration. Clearly the $T_c$ in the interlayer
tunneling mechanism falls slower than usual Josephson coupled
superconductors for low concentration of impurity, but at
larger concentrations it falls very steeply to zero. The critical concentration
of impurity is much smaller in the interlayer mechanism.

In conclusion , we have pointed out that even though Josephson coupling
between plane increases $T_c$, the single particle hopping between the
planes reduce $T_c$. For larger values of $t_{\perp}$ , the increase of
$T_c$ by Josephson tunneling is taken over by the single particle
hopping between the planes at any finite temperatures, and $T_c$ will
decrease with increase of $t_{\perp}$.
Next we considered the effect on $T_c$ by magnetic impurity substitution
out of the plane, where the magnetic moment does not have any direct
exchange coupling with the conduction electrons in the plane, and it
does not change the inplane electronic parameters appreciably like $Pr$ doping
at $Y$ sites does in the $YBCO$ compounds. In the case of purely planar
models, there should not be any suppresion of $T_c$, but with a non zero
effective band term along the $c$ axis, superconductivity will be
suppresed due to both by spin flip scattering by the moment as well
as due to phase slip processes coming from the travelling cooper
pairs along the $c$ direction.
 For the WHA mechanism, only the second process is operative.
 We have done a quantitative prediction that, for small impurity
 concentration, the fall of $T_c$ with impurity concentration in
 conventional planar superconductors is faster than in WHA case. At
 larger concentration of impurity , on the other hand $T_c$ falls very
 sharply in WHA mechanism. This is so, because in the WHA mechanism
 even though the band motion of single quasiparticle motion along
 $c$ axis is prevented, and hence the first channel of $T_c$ reduction
 process is absent, but due to its peculiar momentum conserving
 nature of pair tunneling the $T_c$ is a very sensitive function
 of the pair tunneling amplitude.

\end{document}